\documentclass[useAMS,usenatbib]{mn2e}

\usepackage[latin1]{inputenc}
\usepackage{graphicx}
\usepackage{epstopdf}

\usepackage{rotate}
\usepackage{amsmath, float}
\usepackage{amssymb}
\usepackage{soul}
\usepackage{txfonts}
\usepackage{fleqn}
\usepackage{color}
\usepackage{comment}

\usepackage{mathrsfs}  
\usepackage{booktabs}
\usepackage{colortbl} 

\newcolumntype{G}{>{\columncolor[gray]{0.8}}l} 

\newcommand{\be}{\begin{equation}}
\newcommand{\ee}{\end{equation}}
\newcommand{\bdm}{\begin{displaymath}}
\newcommand{\edm}{\end{displaymath}}
\newcommand{\bea}{\begin{multline}}
\newcommand{\eea}{\end{multline}}
\newcommand{\ba}{\begin{align}}
\newcommand{\ea}{\end{align}}

\newcommand{\thi}{\theta_{\rm i} }

\def\simlt{\mathrel{\hbox{\rlap{\hbox{\lower4pt\hbox{$\sim$}}}\hbox{$<$}}}}
\def\simgt{\mathrel{\hbox{\rlap{\hbox{\lower4pt\hbox{$\sim$}}}\hbox{$>$}}}}

\title[Particle Escape from Bow-Shock PWNe]{Escape of High Energy
  Particles from Bow-Shock Pulsar Wind Nebulae}
\author[N. Bucciantini]{
N. Bucciantini$^{1,2,3}$\thanks{E-mail: niccolo@arcetri.astro.it} \\
$^{1}$INAF - Osservatorio Astrofisico di Arcetri, Largo E. Fermi 5,
I-50125 Firenze, Italy\\
$^{2}$Dipartimento di Fisica e Astronomia, Universit\`a degli Studi di Firenze, Via G. Sansone 1, 
I-50019 Sesto F.~no  (Firenze), Italy\\
$^{3}$INFN - Sezione di Firenze, Via G. Sansone 1, I-50019 Sesto F.~no  (Firenze), Italy}

\begin{document}
 
\date{Accepted / Received}

\maketitle

\label{firstpage}

\begin{abstract}
The detection of bright X-ray features and large TeV halos around old
pulsars that have escaped their parent Supernova Remnants and are
interacting directly with the ISM, suggest that high energy
particles, more likely high energy pairs, can escape from these
systems, and that this escape if far more complex than a simple
diffusive model can predict. Here we present for the first time a
detailed analysis of how  high energy particles escape from the
head of the bow shock. In particular we focus our attention on the
role of the magnetic field geometry, and the inclination of the pulsar
spin axis with respect to the direction of the pulsar kick
velocity. We show that asymmetries in the escape pattern of charged
particles are common, and they are strongly energy dependent. More interestingly we show that the flow of
particles from bow-shock pulsar wind nebulae is likely to be charge
separated, which might have profound consequences on the way such flow
interacts with the ISM magnetic field, driving local turbulence.

\end{abstract}


\section{Introduction}
\label{sec:intro}

Once escaped from their parent Supernova Remnant (SNR), usually a few
tens of thousands of years after the Supernova explosion, pulsars
(PSRs) begin to interact directly with the ISM \citep{Gaensler_Slane06a,Bucciantini_Bandiera01a}. The ram pressure
balance between the pulsar wind and the supersonic ISM flow (in the
reference frame of the PSR itself) leads to the formation of a
cometary nebula known as bow-shock pulsar wind nebula (BSPWN) \citep{Wilkin96a,Bucciantini_Bandiera01a,Bucciantini02b}. These
nebulae have been observed in H$_{\alpha}$ emission, due to neutral Hydrogen
of the ISM interacting through charge-exchange with the ionized
component of the ISM itself, shocked in the outer
bow-shock \citep{Kulkarni_Hester88a,Cordes_Romani+93a,Bell_Bailes+95a,van-Kerkwijk_Kulkarni01a,Jones_Stappers+02a,Brownsberger_Romani14a,Romani_Slane+17a}, and in non-thermal radio/X-rays associated with the shocked
pulsar wind, flowing backward into long tails \citep{Arzoumanian_Cordes+04a,Kargaltsev_Pavlov+17a,Kargaltsev_Misanovic+08a,Gaensler_van-der-Swaluw+04a,Yusef-Zadeh_Gaensler05a,Li_Lu+05a,Gaensler05a,Chatterjee_Gaensler+05a,Ng_Camilo+09a,Hales_Gaensler+09a,Ng_Gaensler+10a,De-Luca_Marelli+11a,Marelli_De-Luca+13a,Jakobsen_Tomsick+14a,Misanovic_Pavlov+08a,Posselt_Pavlov+17a,Klingler_Rangelov+16a,Ng_Bucciantini+12a}.

Modeling of BSPWNe have progressed in the last decades from
simple classical axisymmetric  hydrodynamics \citep{Bucciantini02b} to
relativistic MHD \citep{Bucciantini_Amato+05a} and full and/or
simplified 3D \citep{Vigelius_Melatos+07a,Bucciantini18b,Barkov_Lyutikov18a}. All of those models
are based on the assumption that the pulsar wind behaves as a fluid, and
as such is fully confined within a contact discontinuity (CD)
bounding the non-thermal tail. However, old pulsars are often found
embedded into extended TeV halos \citep{Gallant07a,Helfand_Gotthelf+07a,H.E.S.S.Collaboration_Abramowski+14a,H.E.S.S.Collaboration_Abdalla+18a,Abeysekara_Albert+17a}, that do not resemble at
all the cometary structures found in simulations. It has been
suggested that those PSRs might be slowly moving and still confined
into their parent SNR. However more recently an extended TeV halo was
found surrounding Geminga \citep{Abeysekara_Albert+17a}, which is a well known BSPWN
\citep{Posselt_Pavlov+17a}, where X-rays clearly show a cometary shape. On the other
hand, in the Guitar nebula associated with PSR B2224+65 \citep{Hui_Becker07a} and in
the Lighthouse nebula associated to PSR J1101-6103 \citep{Pavan_Bordas+14a}, bright X-ray
features clearly protruding out of the system, almost orthogonal to
the PSR motion, are observed, that contrast strongly with what fluid
models predict. 

Previous studies have attempted to explain the presence of one-sided
X-ray features invoking a preferential collimated escape of high
energy particles from the PWN \citep{Bandiera08a}. Such directed
outflows have been associated with localized reconnection at the
magnetopause (the CD) between the magnetic field line of the PWN, and
those of the ISM, in a fashion not dissimilar to what is known to
happen on the dayside Earth magnetopause, where it interacts with the
solar wind
\citep{Scholer03a,Frey_Phan+03a,Faganello_Califano+12a,Fuselier_Trattner+12a,Fuselier_Frahm+14b}. However
reconnection at the magnetopause is known to be patchy and sporadic
and to lead only to marginal flux transfer \citep{Kan88a,Pinnock_Rodger+95a,Fear_Trenchi+17a} in contrast
with the persistence of those X-ray features, and on the large
energetics (comparable to the PSR spindown) required to power
them. Moreover, reconnection is likely to affect only those low energy
particles that are bound to follow magnetic field lines, while
existing models invoke the presence of particles
with typical Larmor radii comparable with the size of the bow-shock,
that are likely unable to feel the existence of small reconnecting regions.

The problem of the escape of particles from old BSPWNe, is also
relevant in the context of dark matter searches. Pulsars are likely
the most efficient antimatter factories in the Galaxy, and BSPWNe have
could be one of the major, if not the main, contributor to the positron excess
observed by PAMELA
\citep{Adriani_Barbarino+09a,Hooper_Blasi+09a,Blasi_Amato11a,Adriani_Bazilevskaya+13a,Aguilar_Alberti+13a},
in competition with dark matter  \citep{Wang_Pun+06a}. 

Of course the flow dynamics in these systems can be quite complex, and
many key factors (turbulence, differential
acceleration at the termination shock, etc...) can lead to
anisotropies in the escape of high energy particles. Here, however, our interest
is focused in singling out the specific role of the magnetic field
geometry, using a simplified model for the magnetic field structure in
the head of these nebulae. We trace the trajectories of charged
particles that emerge out of the PWN in its head, and assess the
level of anisotropy associated to the emerging flow.

In Sect.~\ref{sec:traj} we describe how we model the magnetic field
structure and particle trajectories. In Sect.~\ref{sec:res} we
illustrate our results for different magnetic field configurations and
particle rigidities. In Sect.~\ref{sec:conc} we state our conclusions. 

\section{Modeling Particles Trajectories}
\label{sec:traj}

We are interested here in understanding the level of anisotropy that
the magnetic field can induce in the escape of high energy particles
from the head of a typical BSPWN. Given that only particles with
typical energies close to the pulsar voltage are expected to escape
efficiently from the head of these nebulae \citep{Bandiera08a}, and that for these
particles, the Larmor radius in the typical ISM magnetic field is
much larger than the size of the bow shock, only the PSR wind magnetic
field compressed in the head of these nebulae can in principle
introduce anisotropies in their escape. For this reason we only model
the magnetic field inside the bow-shock CD.

The magnetic field structure and the
flow geometry are modelled according to the recipe described in
\citet{Bucciantini18b}, to which the reader is referred for a detailed
discussion. Here we remind briefly their properties. The flow speed is
assumed to be purely laminar, and time independent. Within the PSR
wind termination shock (TS)
the flow is purely radial and moves at the speed of light $c$. In the region
bounded by the TS, on the inner side, and the CD, on the outer side, we use a semi-analytical
prescription, that approximates the average flow conditions, as found in 2D
numerical simulations \citep{Bucciantini02b,Bucciantini_Amato+05a}. The flow is assumed to be purely
axisymmetric, with the symmetry axis given by the direction defined by
the PSR kick velocity with respect to the ISM. On top of this flow we compute the magnetic field
structure, evolving the induction equation for a passive solenoidal field in the steady
state regime, assuming that in the PSR wind the field is purely
toroidal. This is done for various inclinations $\thi$ [see Fig. 3 of \citet{Bucciantini18b}]  between the PSR spin
axis and the PSR velocity, and for different dependencies of the
magnetic field strength on the polar angle $\Psi$ with respect to the PSR
spin axis itself. In the PSR wind, whence the field is injected, the value of the magnetic
field is:
\begin{align}
B = B_o\left(\frac{d_o}{2r}\right)\times\begin{cases}
1-2\Theta[\pi/2-\Psi]\quad{\rm Case \;A}\\
\sin{(\Psi)}(1-2\Theta[\pi/2-\Psi])\quad{\rm Case \;B}\\
\sin{(\Psi)}{\text{tanh}}(\pi/2-\Psi)\quad{\rm Case\; C}\\
\end{cases}
\end{align}
where $r$ is the distance from the PSR (which we locate at the
center of our coordinate system), $\Theta$ is the Heaviside function
that ensures the correct reversal of magnetic polarity at the PSR wind
equator, and $d_o$ the bow-shock stand-off distance (the distance from
the PSR of the CD in the head of the bow-shock) defined as:
\begin{align}
d_o=\sqrt{\frac{L}{4\pi c\rho_o V^2}}
\end{align}
where $L$ is the pulsar spin-down luminosity, $\rho_o$ is the ISM
density (the density of the ionized component), and $V$ is the relative speed of the PSR with respect to
the ISM. $B_o$ is the reference strength of the wind magnetic field at a distance
from the PSR  equal to half the stand-off distance. Case A
represents a field with uniform strength, and a uniform magnetic energy
flux in the wind. Case B instead corresponds to a
magnetic energy flux that scales as  $\sin^2{(\Psi)}$ a prescription
often adopted in PWNe numerical simulations \citep{Komissarov_Lyubarsky04a,Del-Zanna_Amato+04a,Bogovalov_Chechetkin+05a,Volpi_Del-Zanna+08a,Porth_Komissarov+14a,Olmi_Del-Zanna+16a,Del-Zanna_Olmi17a} and based on force free
models of the pulsar magnetospheres \citep{Contopoulos_Kazanas+99a,Timokhin06a,Spitkovsky06a}. Case C adds a further
modulation that suppresses the magnetization in a large sector around
the pulsar equator, and corresponds to the case of a striped wind
emanating from an oblique rotator, where the magnetic field of the
stripes is dissipated \citep{Lyubarsky_Kirk01a,Kirk_Skjaeraasen03a,Lyubarsky03a,Komissarov_Lyubarsky04a,Del-Zanna_Amato+04a}. 

On top of the magnetic field structure found in the various
cases, and for  various inclinations $\thi$, we compute the
trajectories of high energy charged particles, integrating the
equations for energy and momentum, under the action of  the total
Lorentz force, where the electric field is given by the Ideal MHD
condition, the flow speed and the magnetic field. Given that the flow
structure and the magnetic field geometry, in our simplified approach,
are independent of the magnetic field strength (the magnetic field is
passive), the trajectories of the particles are just a function of the
effective rigidity which we parametrize as:
\begin{align}
\mathcal{R}=\frac{mc^2\gamma_i}{qB_o}\frac{1}{d_o} = \frac{R_{\rm lo}}{d_o} 
\end{align}
where $m$ is the mass of the particle, $q$ the modulus of its charge, $\gamma_i$ its Lorentz factor at injection. $\mathcal{R}$ is
just the ratio of the Larmor radius that the particle has in a field
of strength $B_o$, with respect to the stand-off distance. Particles
for which $\mathcal{R} \ll 1$ are strongly tied to the flow and
unlikely to efficiently escape, while particles for which
$\mathcal{R}\gg 1$ escape the bow-shock unaffected by the magnetic
field. Particles with $\mathcal{R}=1$ have typical Larmor radii, in
the typical magnetic field of the head, comparable to the stand-off
distance. It can be shown \citep{Bandiera08a}  that the typical
energies of these particles ranges from about 10 TeV, for a pulsar like PSR
B2224+65 in the Guitar nebula, to a few hundreds TeV, for a pulsar like
Geminga, PSR J0633+1746.

Given that we are interested in the possible escape  from
the very head of these nebulae, particles are followed in their
trajectories until they either reach the CD, or the move
backward at a distance exceeding $2d_o$ from the PSR, at which point we
consider them lost in the tail. It is still likely that those particles
can escape from the tail, but given that the tail might be in general
more turbulent than the head we expect the escape there to be more
isotropic.

In principle, the canonical model of PWNe assumes that particles are
accelerated at the wind TS. However, for simplicity,
given that we are interested in evaluating the level of anisotropy
introduced by the magnetic field, we assume all high energy particles
to be injected from the PSR (or equivalently the pulsar wind),
with an isotropic distribution, and in the radial direction.

\section{Results}
\label{sec:res}

Using the magnetic structure computed according to the recipes of
\citep{Bucciantini18b} we injected $10^5$ particles with high Lorentz factor
($v \rightarrow c$), isotropically from the
pulsar, and we followed their trajectories until they reached the CD.
This was done for both signs of charge. Note that the trajectories of
particles of the same energy depend on the sign of their charge with
respect to the magnetic field polarity, which in turn is related to the
inclination angle between the pulsar dipole moment and the pulsar spin
axis. A trajectory of an electron in a toroidal magnetic field
left-handed with respect to the spin axis, is the same as the
trajectory of a positron of the same energy, in a right-handed
field. Given that there is no way of measuring the polarity of the
magnetic field, one cannot attribute a specific pattern of
trajectories to either positrons/protons or electrons.

Our model allows
us to locate the escape position along the CD, to count the fraction
of particles that escape in the head, and also to evaluate the energy
gains/losses they experience as they move within the BSPWN.

We will focus our discussion on three kind of anisotropies: 
\begin{itemize}
\item {\it head-to-tail} anisotropy: due to the fact that the magnetic
  field compression, the flow speed and the thickness of the BSPWN
  between the TS and CD varies substantially between the head and the
  tail;
\item {\it inclination} anisotropy: due to the different geometry of
the magnetic field, and the associated current layers, for different
inclinations of the PSR spin axis with respect to the PSR kick
velocity;
 \item {\it charge} anisotropy: due to the fact that particles of
   different sign behave differently on current layers/lines, in terms of
   confinement and collimation.
\end{itemize}

\subsection{Fully Axisymmetric Case}

\begin{figure}
	\centering
\includegraphics[width=.50\textwidth, bb= 10 20 560 440,clip]{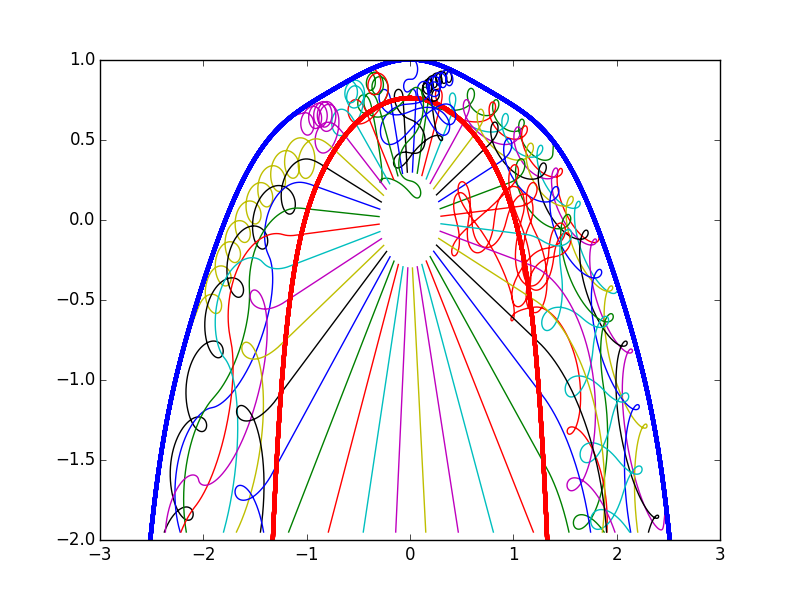}
	\caption{Trajectories of charges particles in the head of a
          BSPWN, for $\mathcal{R}=0.25$. Left part (negative
          abscissae) and right part (positive abscissae)
          represent trajectories of particles of different signs. The
          thick red curve is the PSR wind TS, while the thick blue
          curve is the CD. Axes are in units of the stand off distance
          (the size of the CD in the head).
     }
	\label{fig:track}
\end{figure}

\begin{figure}
	\centering
\includegraphics[width=.50\textwidth, bb= 10 20 650 440,clip]{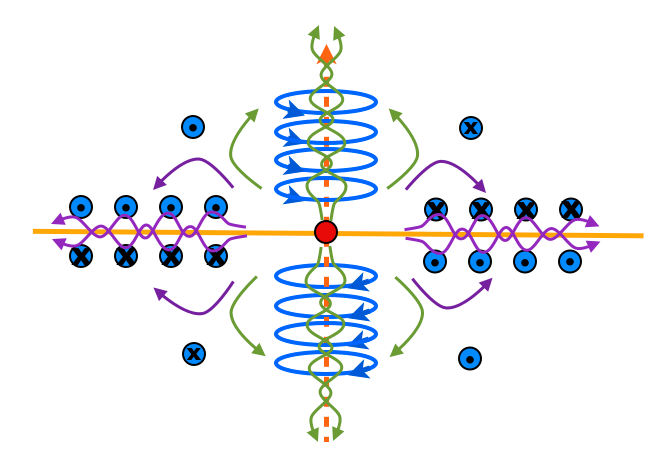}
	\caption{Schematic representation of the trajectories of high
          energy particles in the azimuthal magnetic field,
          injected by the PSR wind (red sphere). The magnetic field, with its
          polarity is shown in blue. The dashed orange line is the
          polar current line, while the orange solid line represents
          the equatorial current sheet. The trajectories of charges of
          different signs are shown in green and purple. 
     }
	\label{fig:colli}
\end{figure}

If $\thi=0^\circ$ the magnetic field structure is fully
axisymmetric, and the magnetic field if fully azimuthal. In this case, the induction electric field is also
purely poloidal, and the same holds for the total Lorentz force. The
trajectories of the particles are bound to be confined in meridional
planes, at fixed azimuthal angle. This implies that there is no {\it
  inclination} anisotropy. In Fig.~\ref{fig:track} we show the
trajectories of particles injected at different polar angles $\Psi$
along a meridional cut of the BSPWN, for $\mathcal{R}=0.25$ and 
different sign of the charge. It is evident that in this regime the
trajectories are strongly dominated by the flow advection, with gyration
playing only a minor role. It is interesting to note that along the
axis, in the head of the nebula, in the region corresponding to the
location of the polar current line (the symmetry axis of the toroidal
magnetic field), while particles of one sign tend
to be confined, the particles of the opposite sign tend to be
expelled. This is a common property of current lines/sheets, where
magnetic field reverses sign. Moreover one can see in the tail the
change in the direction of gyration, associated with the change of the
magnetic field polarity due to the reversal of the field in the
equatorial plane of the PSR rotation. It is evident that a large
fraction of particles are advected in the tail. and only a small
fraction manages to reach the CD in the head. In Fig.~\ref{fig:colli}, we illustrate
schematically how charged particles behave in the presence of current lines/sheets. At least as long as the the magnetic field preserves its
azimuthal geometry, the polar  current lines will confine particles of one
sign, while the equatorial current sheet will confine particles of the
opposite sign. Of course this is an energy dependent statement. In
the limit $\mathcal{R}\rightarrow 0$ particles are tied to the flow,
while for $\mathcal{R}\rightarrow \infty$ particles will not feel
the magnetic field. In general we found that for $\mathcal{R} =3$ the
location of the particle escape points on the CD is almost isotropic,
and the totality of particles that are injected escapes. On the other
hand for $\mathcal{R} < 0.1$ the fraction of particles that escape in
the head drops below $5-10\%$, and the vast majority is advected in the tail.

\begin{figure*}
	\centering
\includegraphics[width=.25\textwidth, bb= 90 20 500 440,clip]{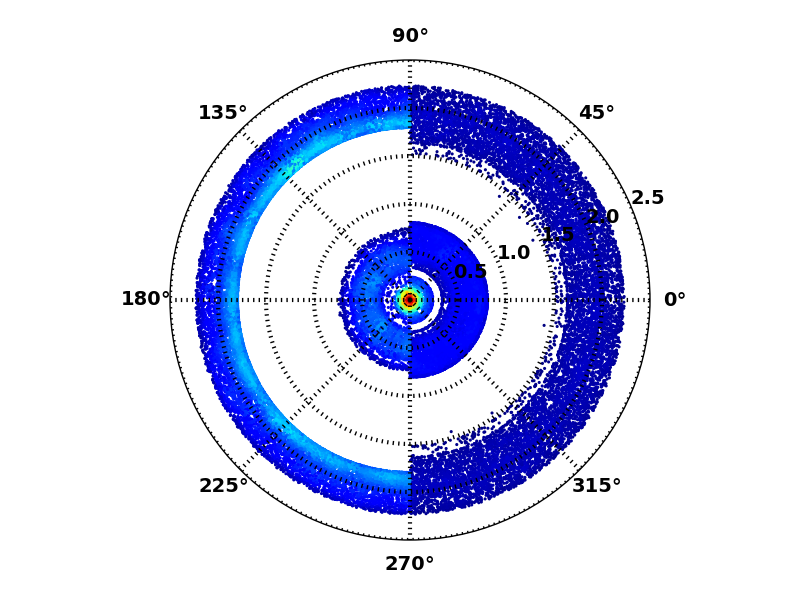}\includegraphics[width=.25\textwidth, bb= 90 20 500 440,clip]{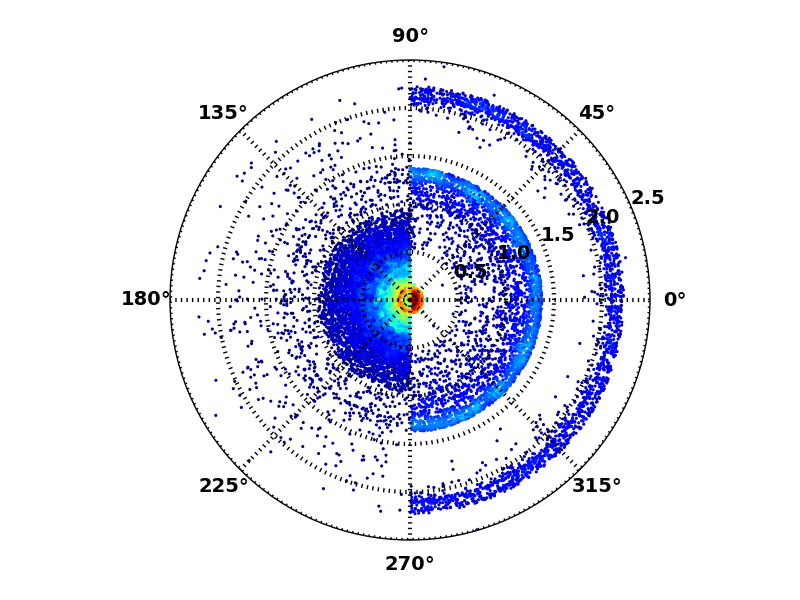}\includegraphics[width=.25\textwidth, bb= 90 20 500 440,clip]{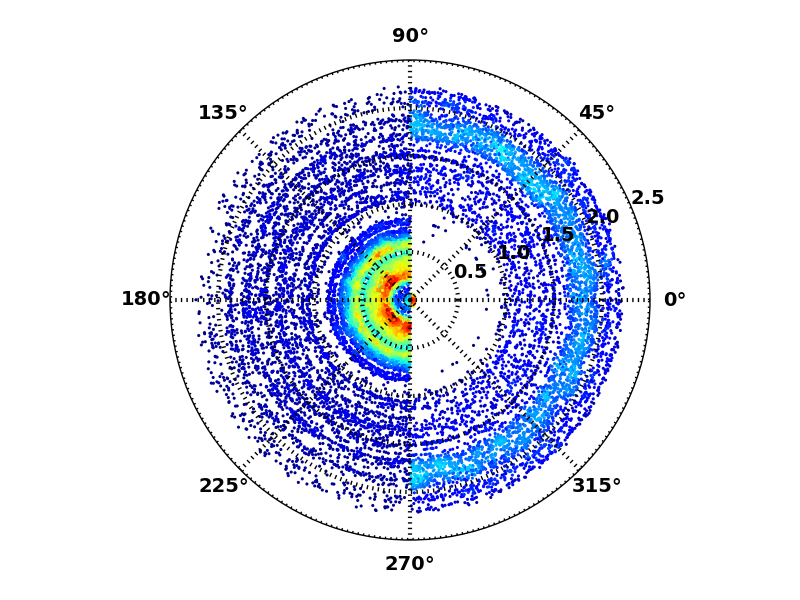}\includegraphics[width=.25\textwidth, bb= 90 20 500 440,clip]{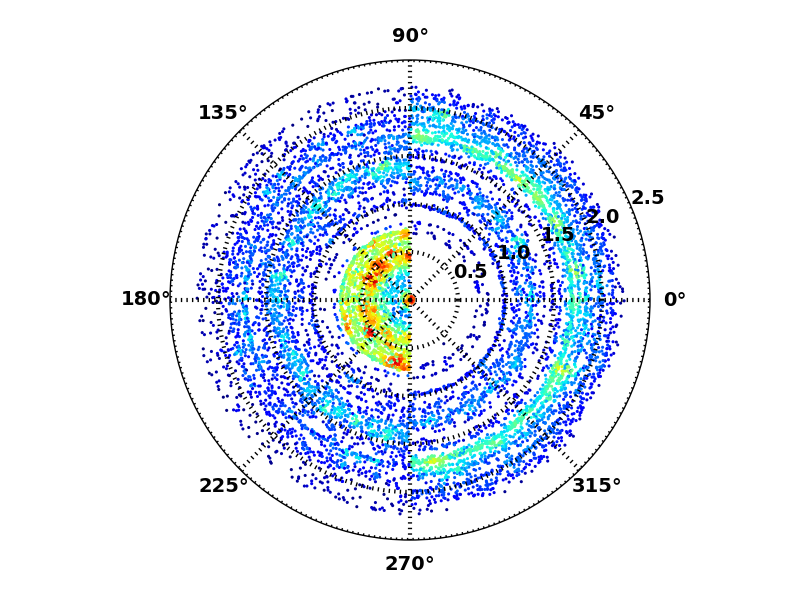}\\
	\includegraphics[width=.25\textwidth, bb= 90 20 500
        430,clip]{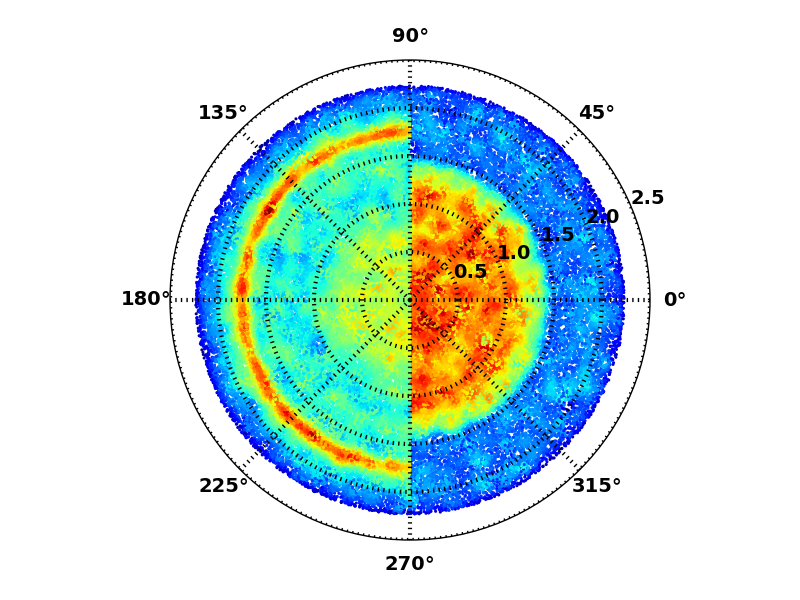}\includegraphics[width=.25\textwidth,
        bb= 90 20 500
        430,clip]{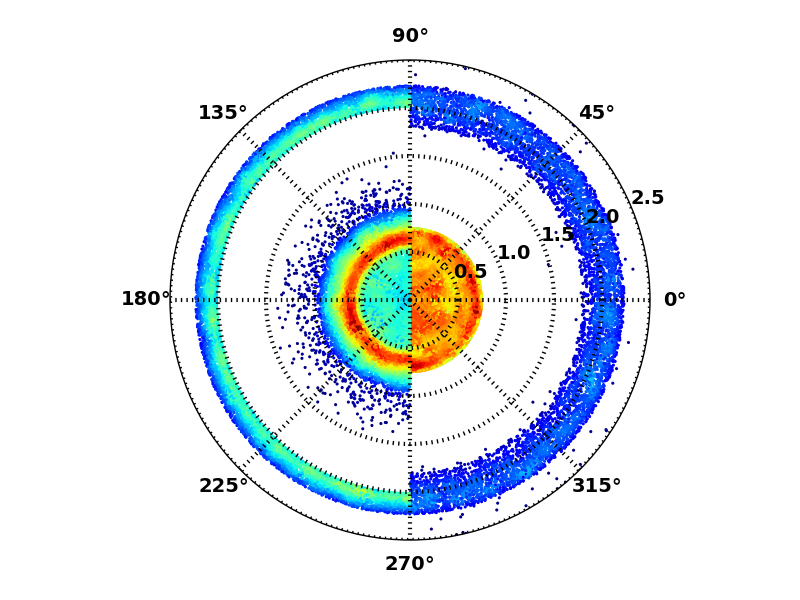}\includegraphics[width=.25\textwidth,
        bb= 90 20 500
        430,clip]{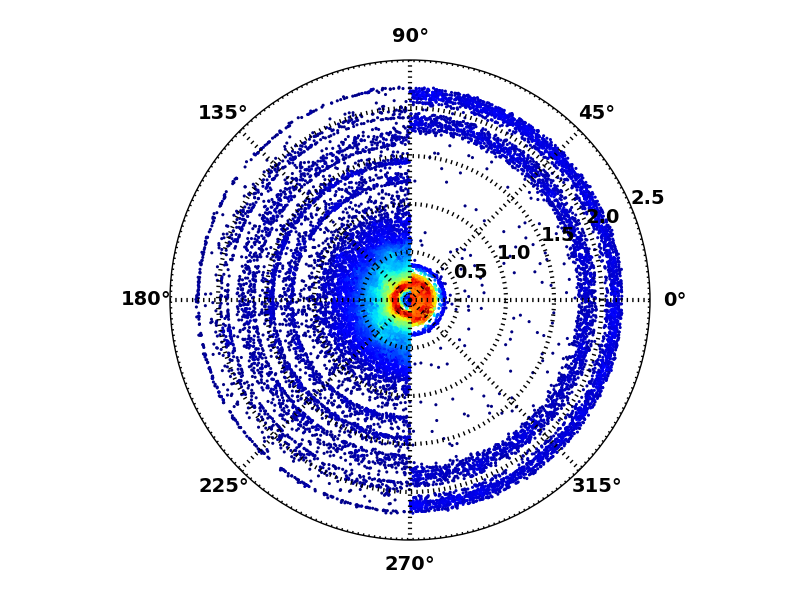}\includegraphics[width=.25\textwidth,
        bb= 90 20 500 430,clip]{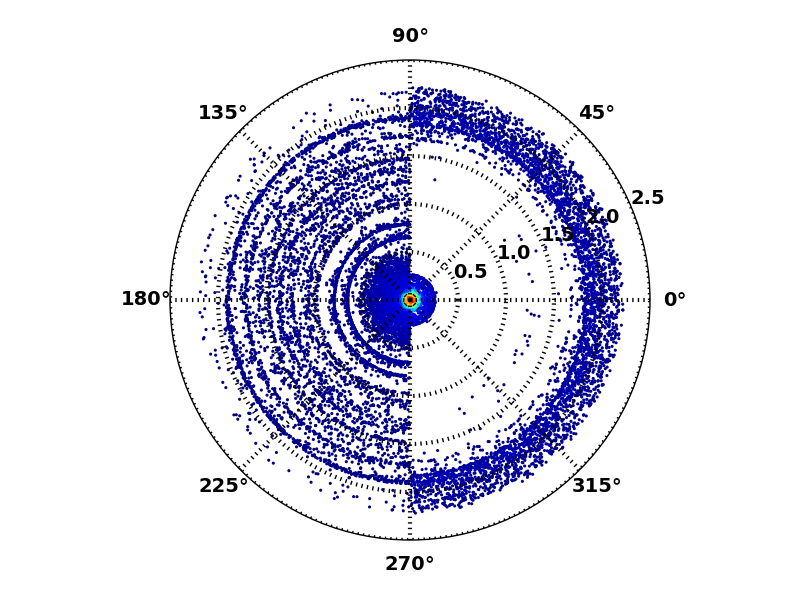}\\
\includegraphics[width=.25\textwidth, bb= 90 20 500
430,clip]{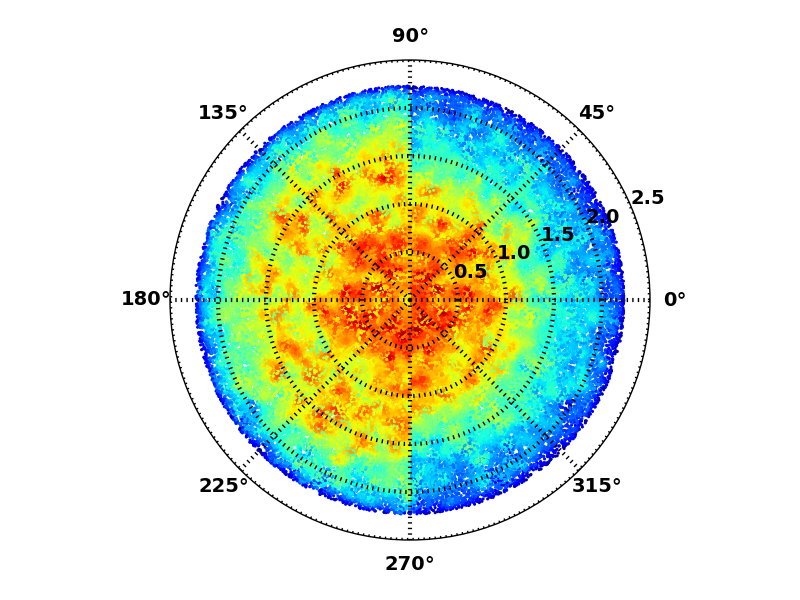}\includegraphics[width=.25\textwidth,
bb= 90 20 500
430,clip]{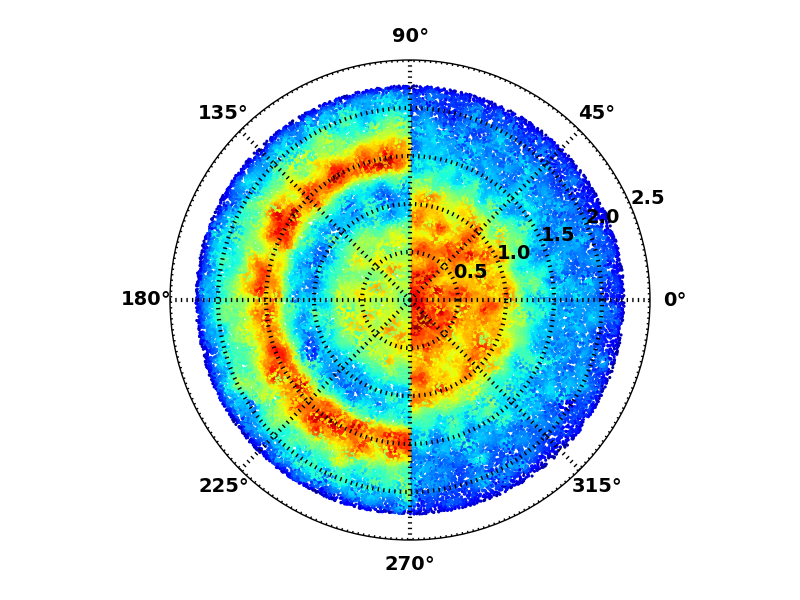}\includegraphics[width=.25\textwidth,
bb= 90 20 500
430,clip]{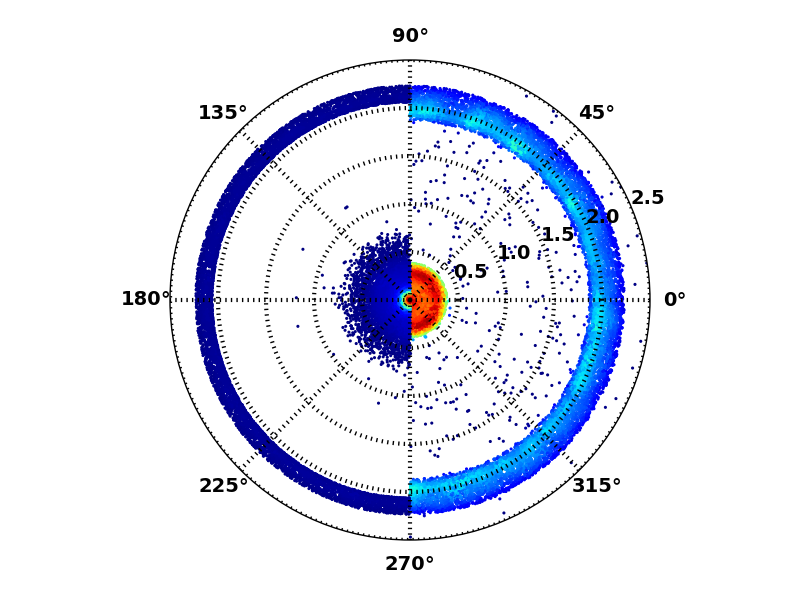}\includegraphics[width=.25\textwidth,
bb= 90 20 500 430,clip]{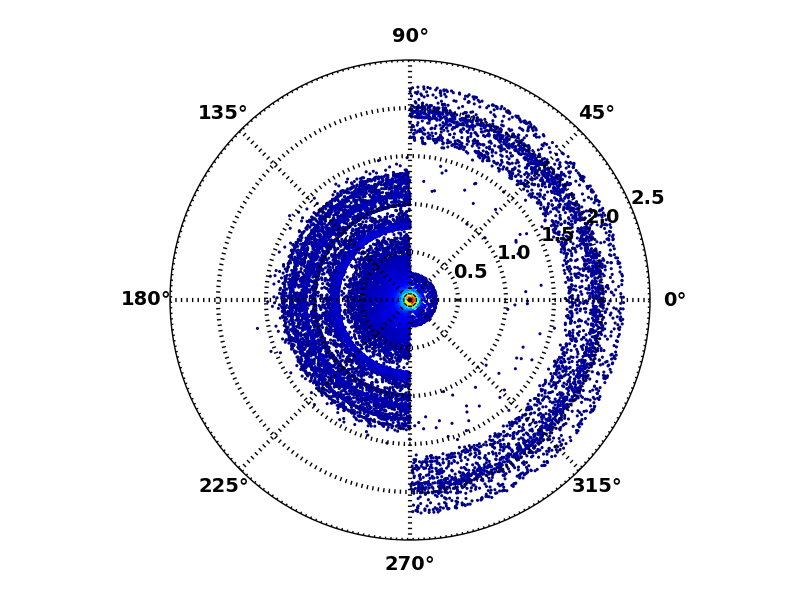}\\
\includegraphics[width=.25\textwidth]{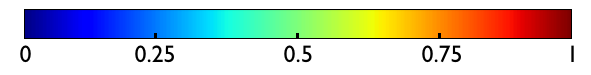}\\
	\caption{Escape location of charged particles along the CD of
          the BSPWN, in polar coordinates for the fully axisymmetric case $\thi=0^\circ$. The polar angle $\theta$ in
          the range $[0,2.5]$rad is measured along the CD from the head
          of the bow-shock (the symmetry axis of the nebula). Given
          the symmetry of the problem, in the left and right side of each panel
          we plot escape positions for particles of different sign in
          the range of 
          azimuthal angle $\phi$ either $[0,\pi]$ or $[\pi,2\pi]$. The
          color scheme ranges from 0 to 1, and represents the
          relative density of points computed with a Gaussian kernel
          of  FWHM $=0.02$ in units of the stand-off distance on the
          projection plane. Rows from top to bottom show cases A, B
          and C. Columns from left to right represent different
          effective rigidities $\mathcal{R}={1,0.5,0.25,0.1}$.    }
	\label{fig:exit00}
\end{figure*}

In Fig.~\ref{fig:exit00} we show the exit points of charged particles
in polar coordinates along the CD, for various values of the effective
rigidity $\mathcal{R}$ and for different latitudinal shapes of the
magnetic field. The appearance of one or more rings is due to the
combination of gyration and advection, that leads to a periodic
pattern. In Case A for small vales of $\mathcal{R}$ one sees that the
escape points, apart from a concentration/depletion in the very head,
due to the presence of the polar current line,  tend to be uniformly
distributed, compatible with the picture of an advection
dominated motion. For higher values of $\mathcal{R}$, once the typical
gyration radius becomes comparable with the size of the system, one begins to see
large modulations in the escape pattern, with regions devoid of
escaping particles. In Case B, one finds that, due to the $\sin$-like
modulation of the magnetic field in the head, the pattern in the head
is changed. For $\mathcal{R}=1$ one begins to see a pattern close to
the free escape limit. In Case C, where a large low-magnetization
equatorial outflow tends to lower the mean magnetic field in the
head, patterns much closer to the free escape limit are already
reached at   $\mathcal{R}=0.5$.  Indeed we find that the escape
fraction for  $\mathcal{R}=0.5$ approaches 50\% of the particle
injected, and values close to 100\% for  $\mathcal{R}=1$.
Interestingly for $\mathcal{R}<1$ the escape fraction of particles of
different sign can differ by a factor up to 2. On the other
hand, the energy gains/losses are only marginal, at most a factor of
2. It is evident, from the large diversity of patterns, that there is a
strong dependence on the energy of the injected particles, on the
specific shape of the CD and TS. However, even if the pattern might
be sensitive to these specific choices of configuration, it is clear that for
$\mathcal{R}<0.5$ large head-to-tail and charge anisotropies might be
present, with charges of different signs escaping at different
locations along the BSPWN.

\subsection{The Orthogonal Case}

\begin{figure*}
	\centering
\includegraphics[width=.25\textwidth, bb= 90 20 500 440,clip]{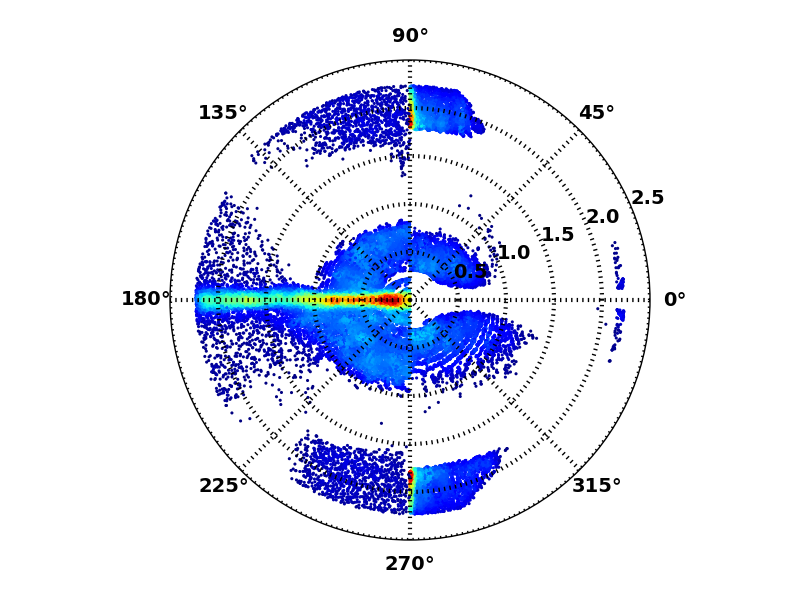}\includegraphics[width=.25\textwidth, bb= 90 20 500 440,clip]{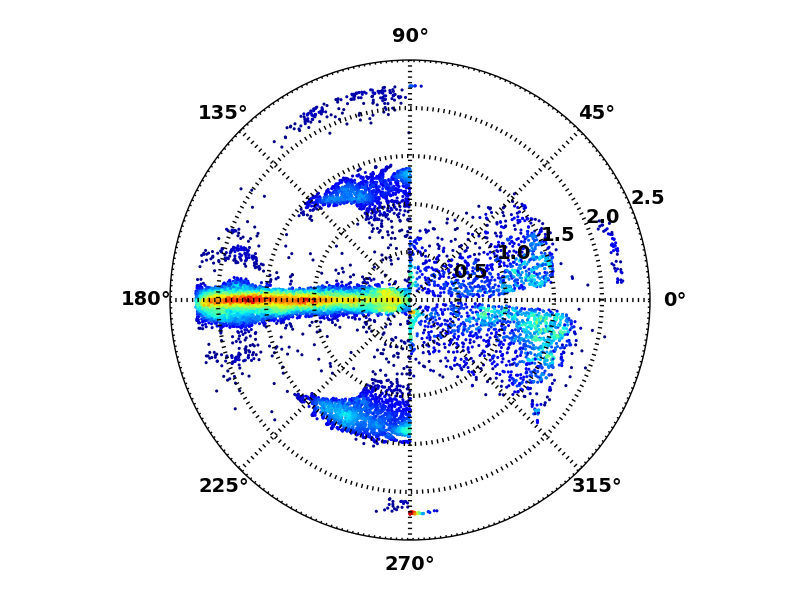}\includegraphics[width=.25\textwidth, bb= 90 20 500 440,clip]{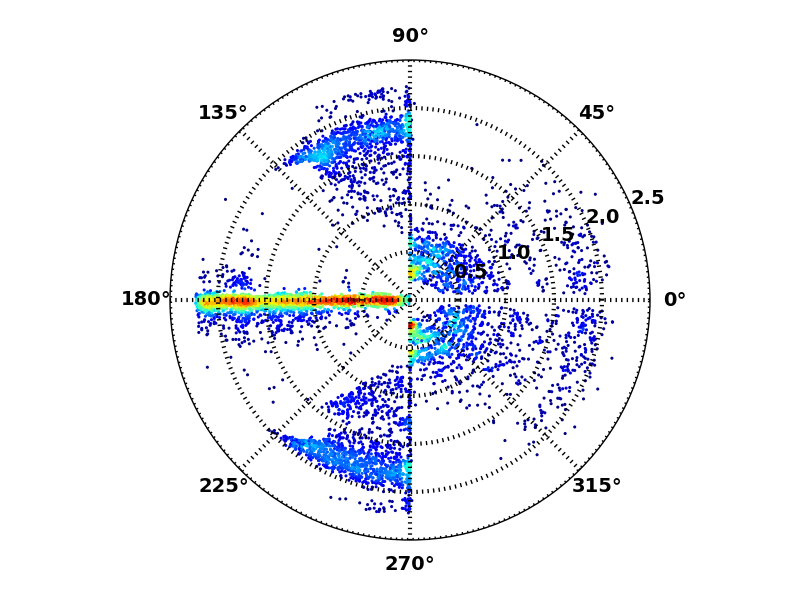}\includegraphics[width=.25\textwidth, bb= 90 20 500 440,clip]{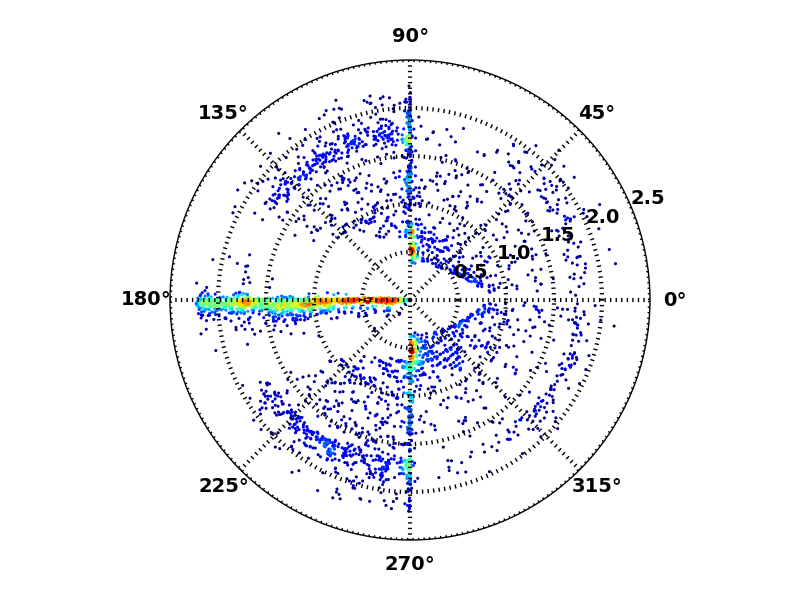}\\
	\includegraphics[width=.25\textwidth, bb= 90 20 500
        430,clip]{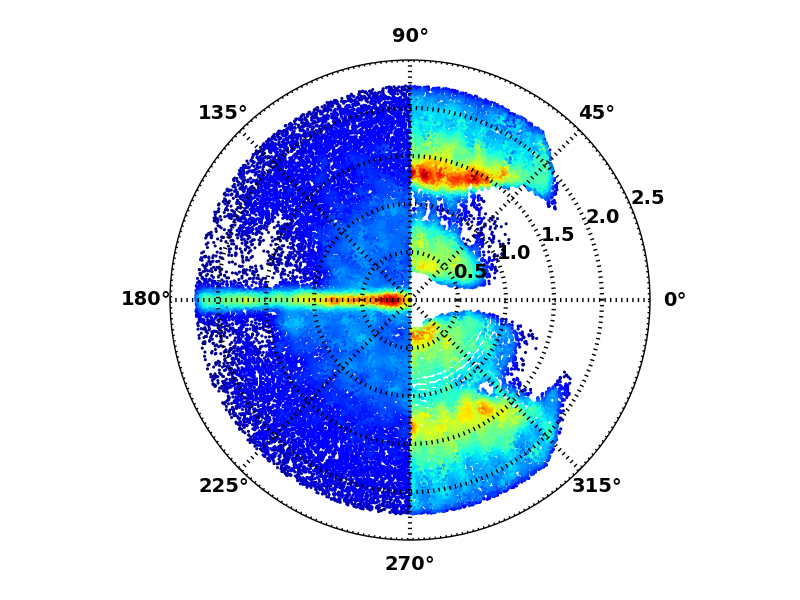}\includegraphics[width=.25\textwidth,
        bb= 90 20 500
        430,clip]{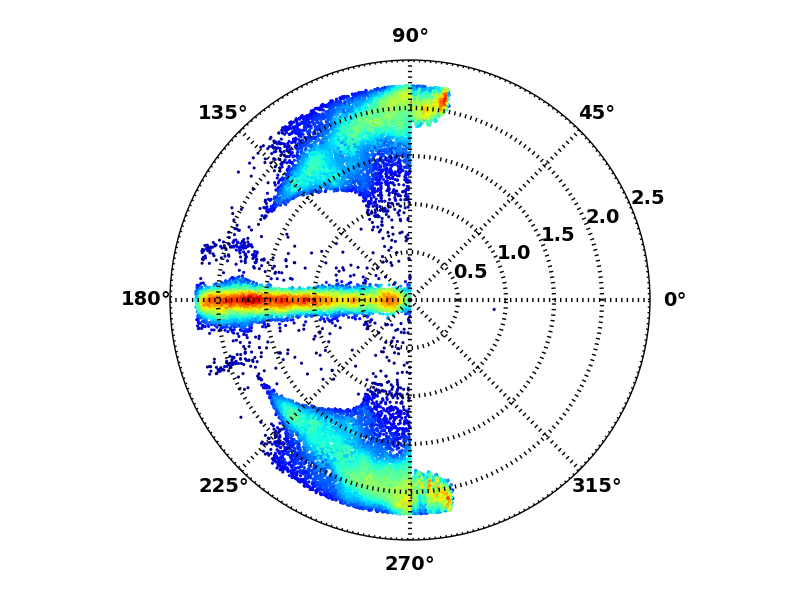}\includegraphics[width=.25\textwidth,
        bb= 90 20 500
        430,clip]{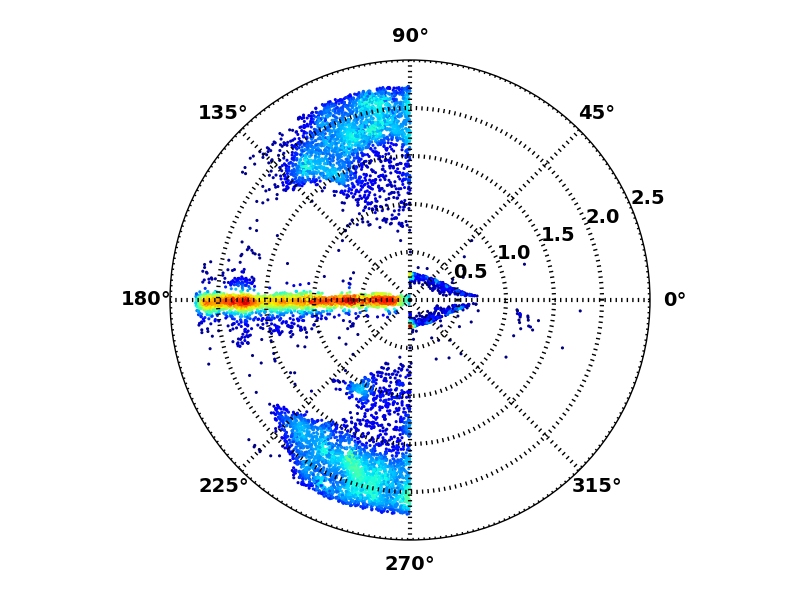}\includegraphics[width=.25\textwidth,
        bb= 90 20 500 430,clip]{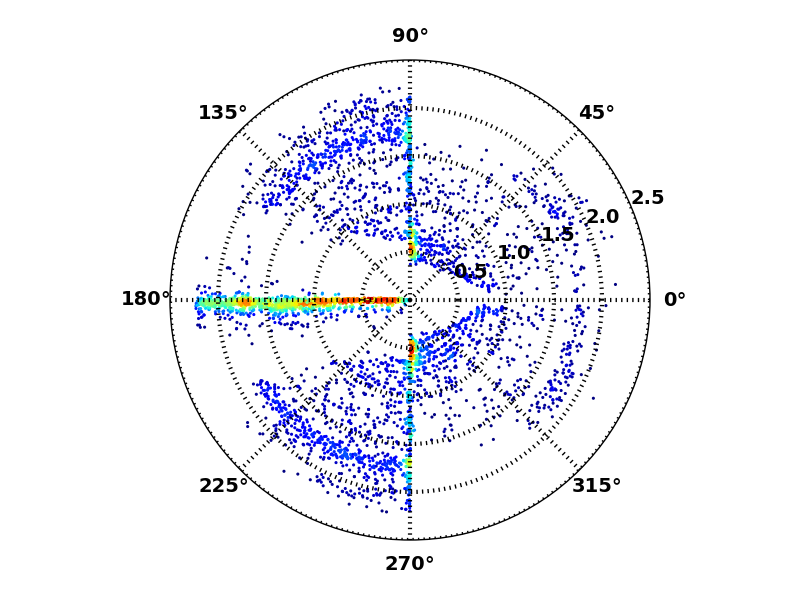}\\
\includegraphics[width=.25\textwidth, bb= 90 20 500 430,clip]{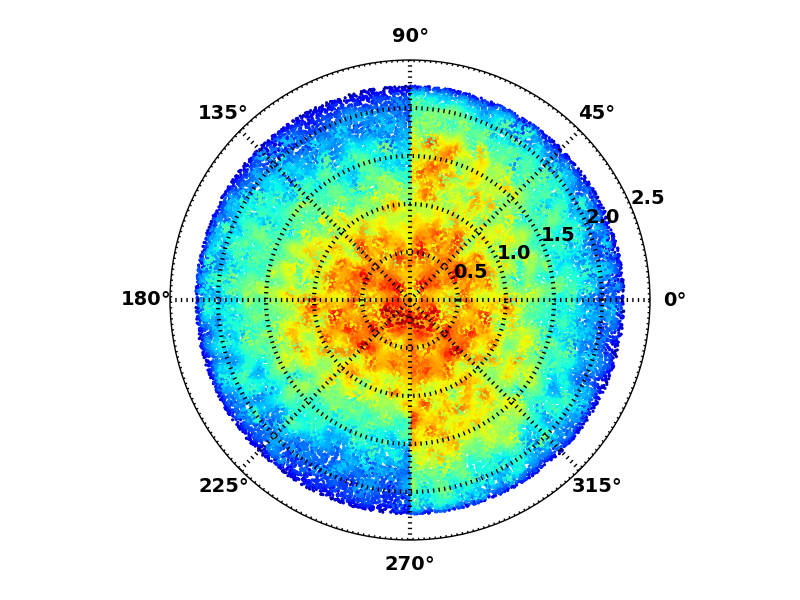}\includegraphics[width=.25\textwidth, bb= 90 20 500 430,clip]{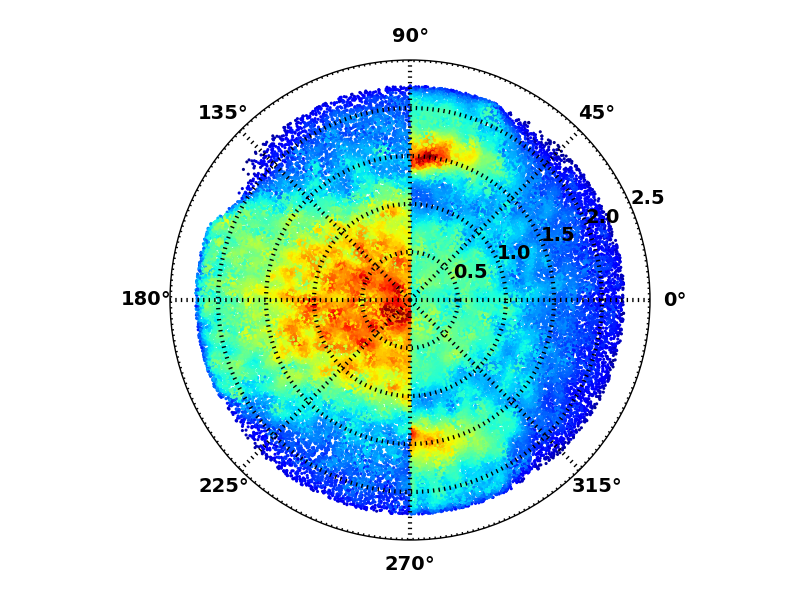}\includegraphics[width=.25\textwidth, bb= 90 20 500 430,clip]{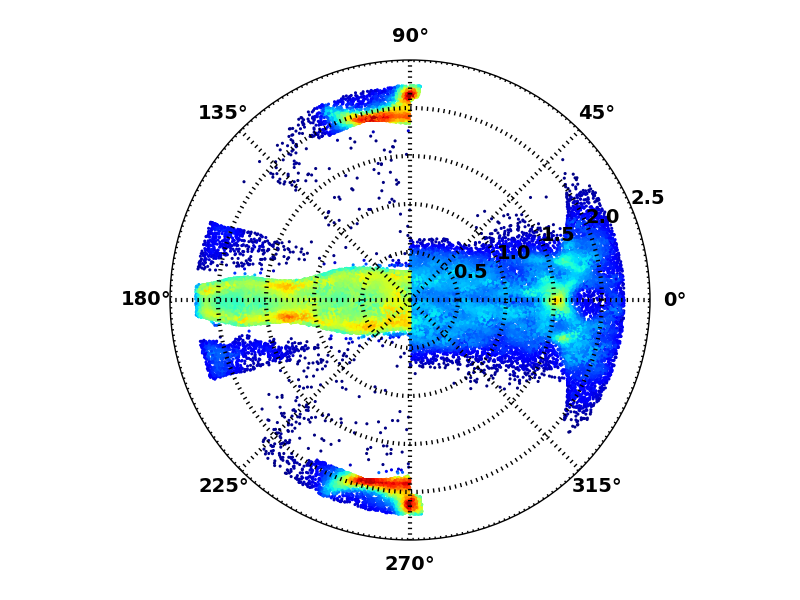}\includegraphics[width=.25\textwidth, bb= 90 20 500 430,clip]{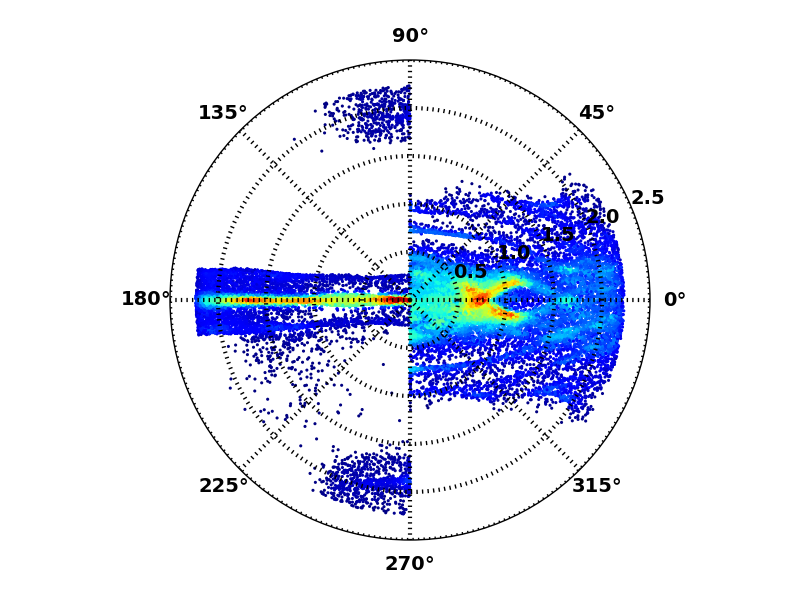}\\
	\caption{Same as Fig.~\ref{fig:exit00} but for the orthogonal
          case $\thi=90^\circ$.}
	\label{fig:exit90}
\end{figure*}

In the orthogonal case, $\thi=90^\circ$, it is the equatorial current
sheet that extends all the way to the CD of the BSPWN, while the polar
current lines remain confined inside the nebula. In this case one can
clearly observe, as shown in Fig.~\ref{fig:exit90}, the charge anisotropy due to the presence of the
equatorial current sheet. While particles of one sign tend to be
strongly concentrated in the current sheet, those of the opposite sign
are strongly defocused. In this case the number of escaping particles
of opposite signs can differ even by a factor 2-3. In cases A and B,
for  $\mathcal{R} < 0.5$ less than 10\% of the particles manage to
reach the CD, away from the curent sheet. For $\mathcal{R} =1 $, in
case B, where the magnetic field close to the polar line is suppressed
in a $\sin{}$-like fashion, one begins to see the escape of particles
confined in the polar region. In case C, where the equatorial current
layer is broad, we clearly see for $\mathcal{R} \ge 0.5$ a pattern
that rapidly approaches the isotropic case, even if a marginal
residual asymmetry between the escape points of the two charges (one more concentrated in
the equatorial plane, one more along the polar axis direction)
remains. Interestingly the escape probability for $\mathcal{R} <1$ in
the orthogonal case if smaller than in the aligned case, suggesting
that the geometry of the field provides a better confinement. More
interesting is the fact that the fraction of escaping particles of
different sign, at the same $\mathcal{R}$ can differ even by a factor
a few. This mostly happens for small $\mathcal{R}<0.5$, and is due to
the confinement effect of the current sheet. Indeed the excess is
almost completely due to particle escaping along the equatorial current
sheet. As before, the energy gains/losses are comparable and not exceeding a
factor a few.

\subsection{Inclined Case}

\begin{figure*}
	\centering
\includegraphics[width=.25\textwidth, bb= 90 20 500 440,clip]{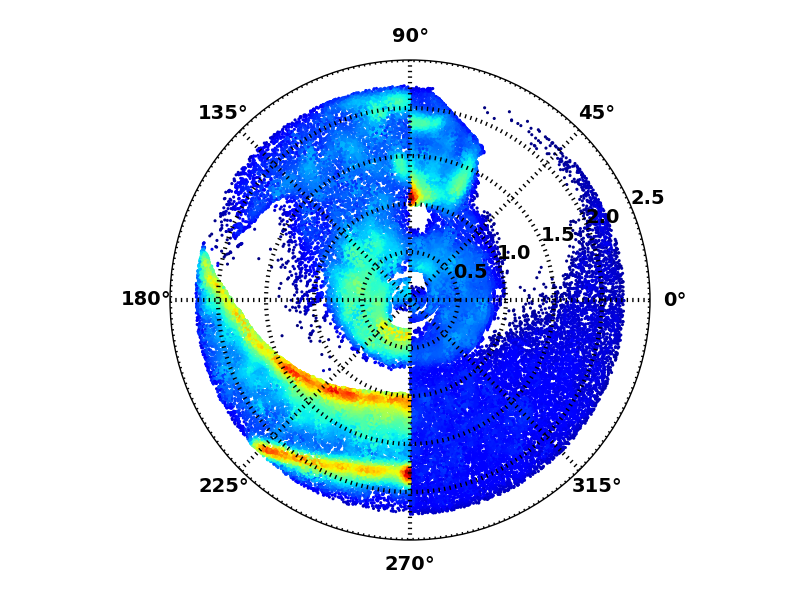}\includegraphics[width=.25\textwidth, bb= 90 20 500 440,clip]{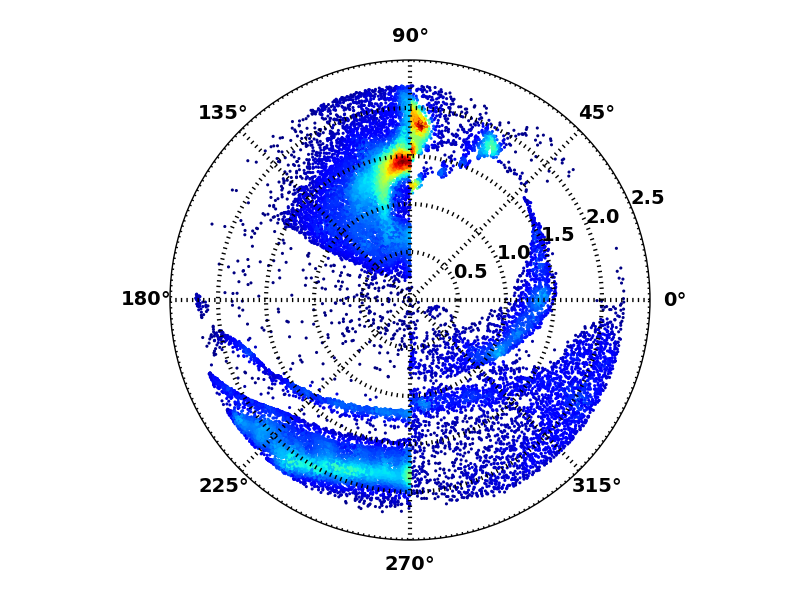}\includegraphics[width=.25\textwidth, bb= 90 20 500 440,clip]{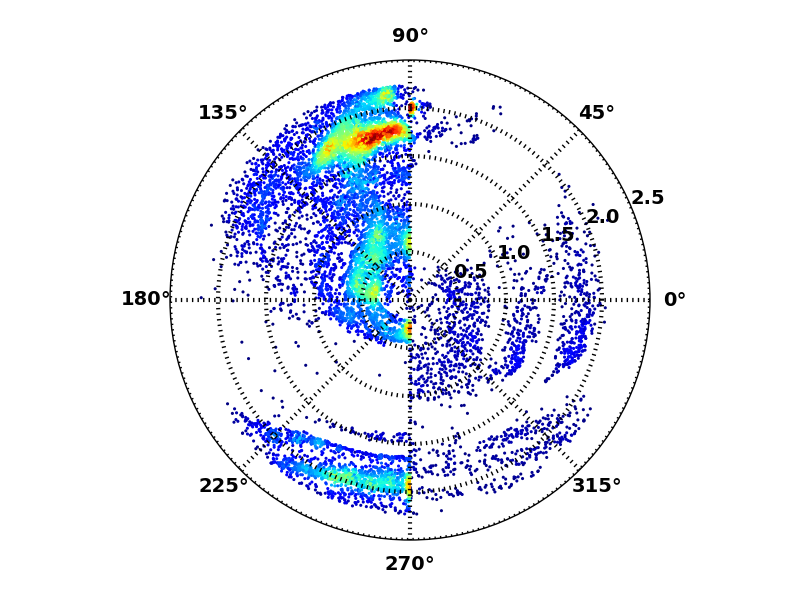}\includegraphics[width=.25\textwidth, bb= 90 20 500 440,clip]{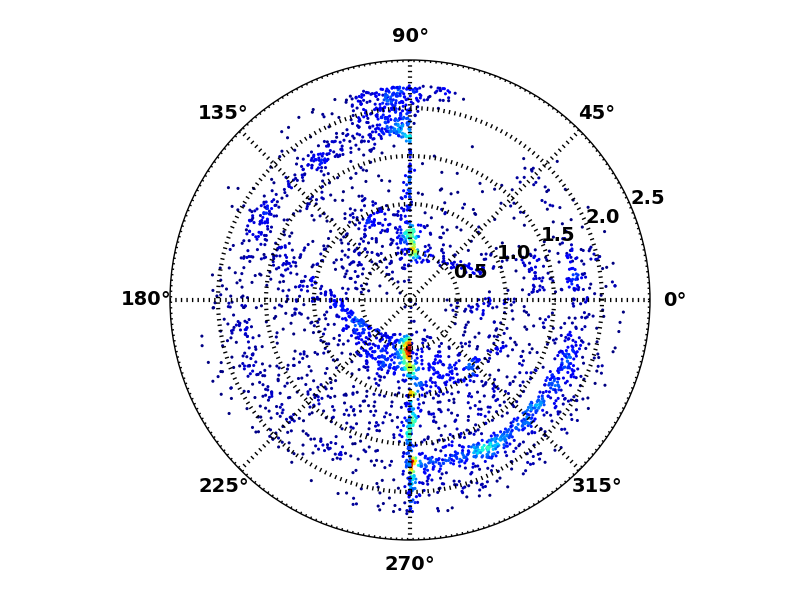}\\
	\includegraphics[width=.25\textwidth, bb= 90 20 500
        430,clip]{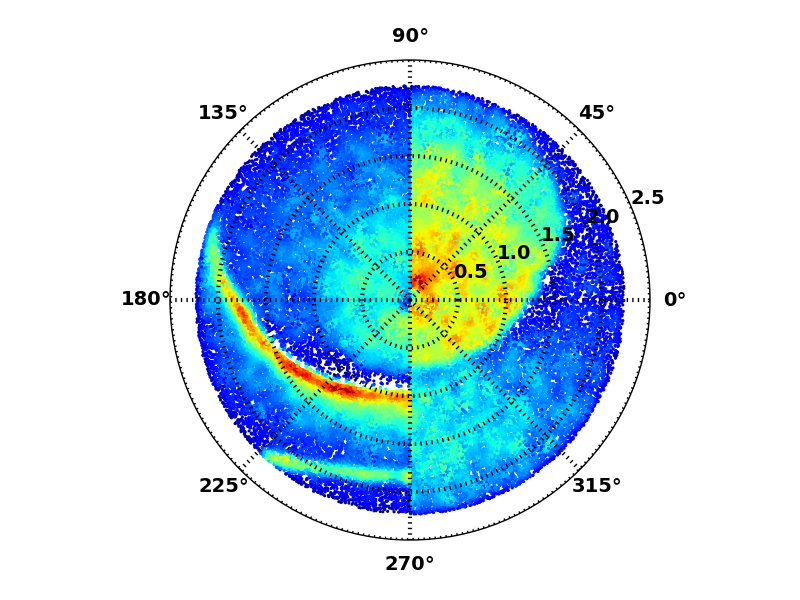}\includegraphics[width=.25\textwidth,
        bb= 90 20 500
        430,clip]{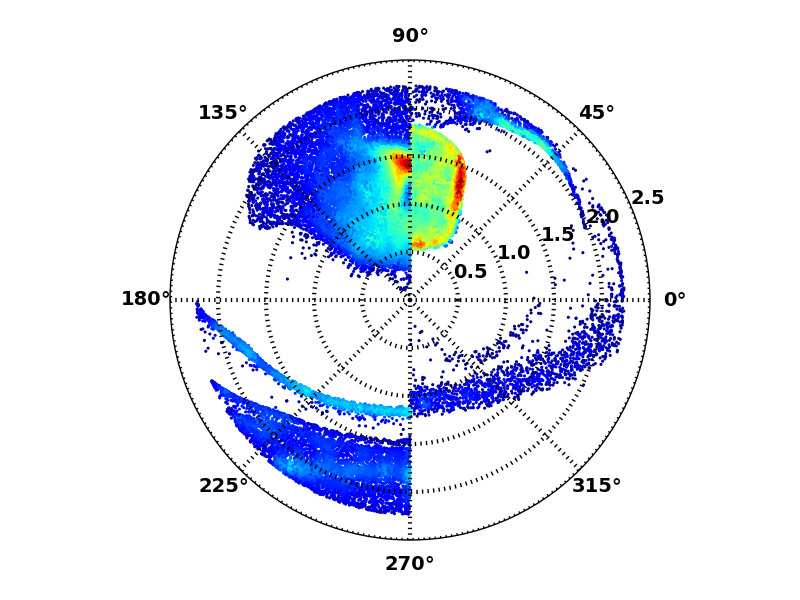}\includegraphics[width=.25\textwidth,
        bb= 90 20 500
        430,clip]{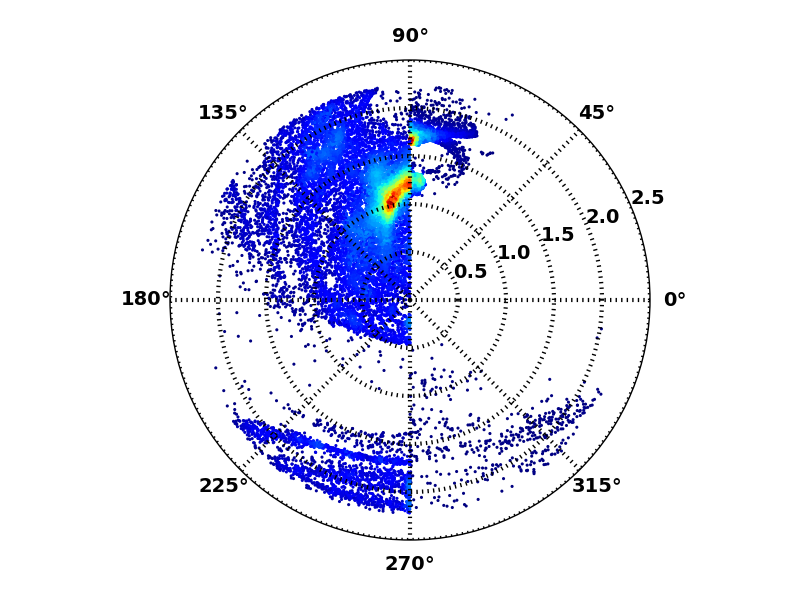}\includegraphics[width=.25\textwidth,
        bb= 90 20 500 430,clip]{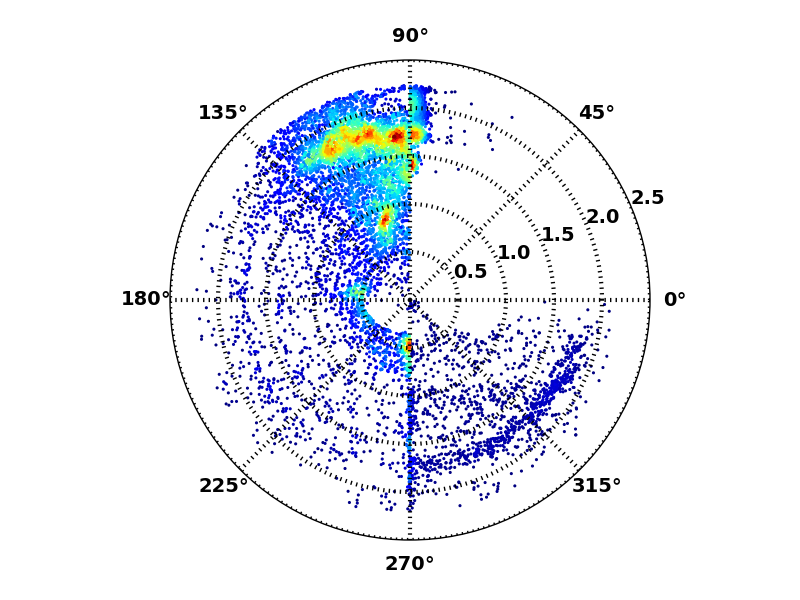}\\
\includegraphics[width=.25\textwidth, bb= 90 20 500 430,clip]{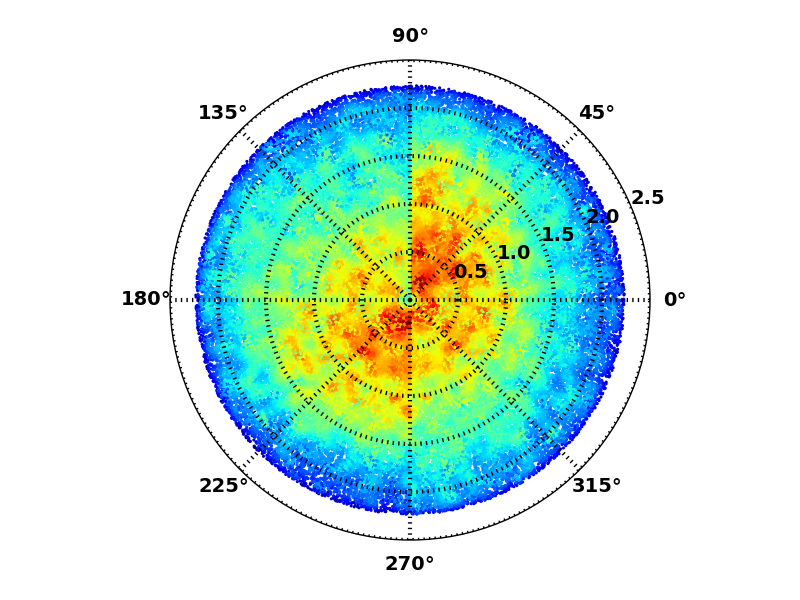}\includegraphics[width=.25\textwidth, bb= 90 20 500 430,clip]{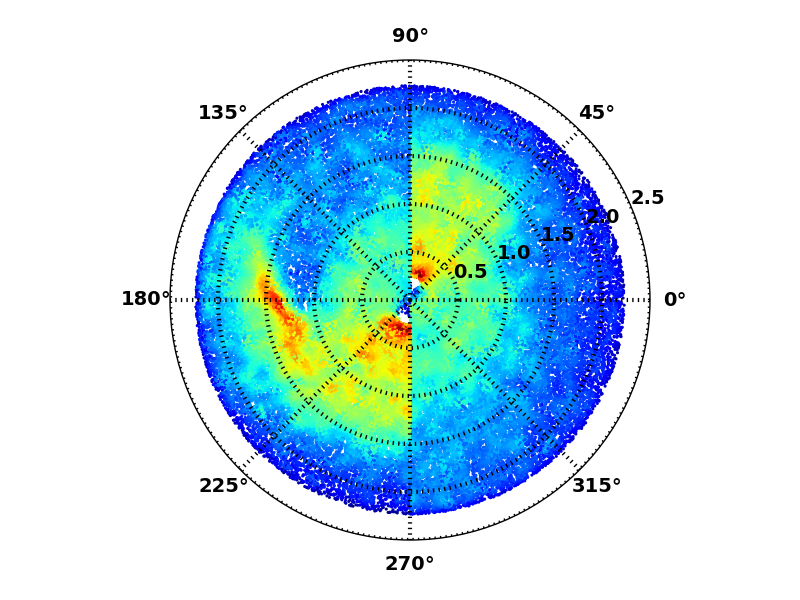}\includegraphics[width=.25\textwidth, bb= 90 20 500 430,clip]{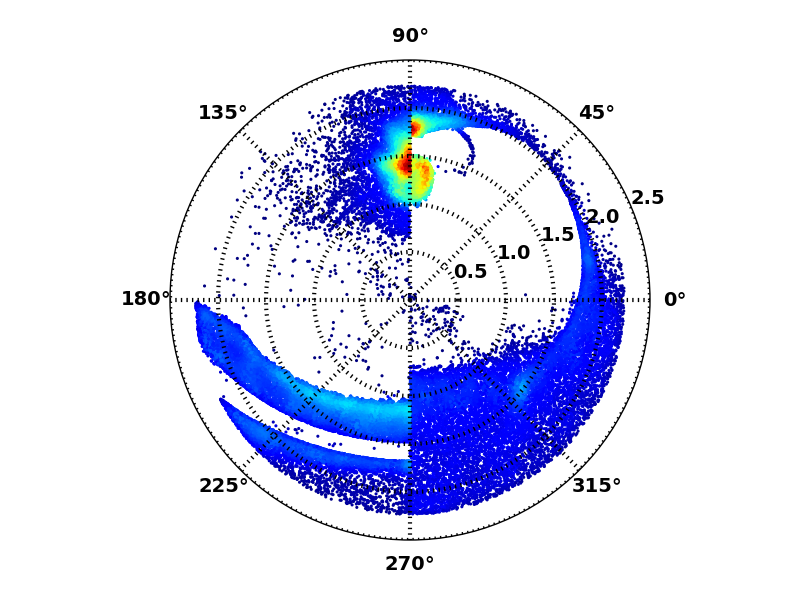}\includegraphics[width=.25\textwidth, bb= 90 20 500 430,clip]{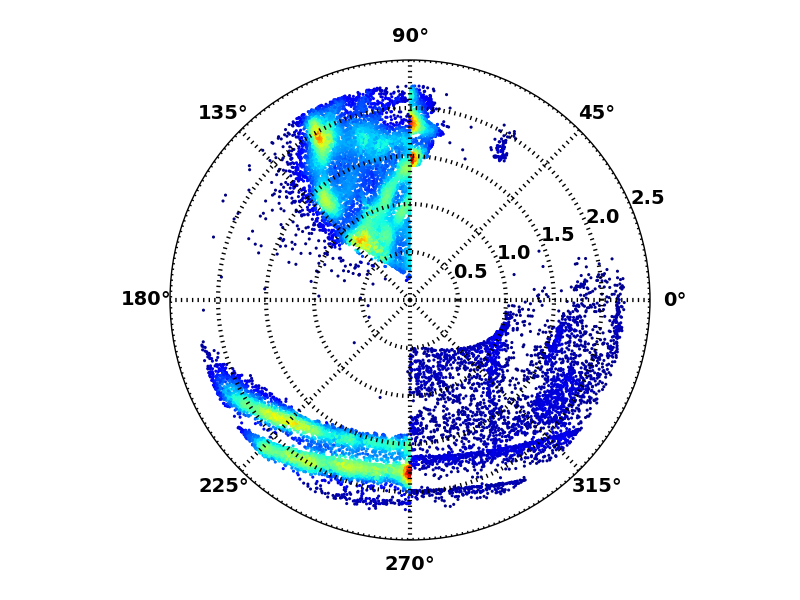}\\
	\caption{Same as Fig.~\ref{fig:exit00} but for the inclined
          case $\thi=45^\circ$.}
	\label{fig:exit45}
\end{figure*}

In the inclined case $\thi=45^\circ$ we see, as shown in Fig.~\ref{fig:exit45}, the same pattern due to
the charge confinement in the equatorial current sheet, which
translates into a clear asymmetry between the upper and lower part of
the CD. Again we find a marked difference in the number of escaping
particles between the two different charges, with escape fraction
smaller than a few percent for $\mathcal{R} < 0.25$. The presence of
the equatorial current sheet, which now extends to the bottom part of
the CD, and of a polar current line in the upper part becomes evident at
$\mathcal{R} >0.5$ in the escape pattern of the opposite charges. This
top-bottom asymmetry is still present for $\mathcal{R}=1$ even in case
C where both the polar current line and equatorial current sheet are
broader. Similar to the previous cases there are regions on the CD
where no escaping particle is present. Even in this case, gain and
losses are in general negligible. As in the orthogonal case, even here
we do observe a marked difference in the number of escaping particles
of different signs.

\section{Conclusions}
\label{sec:conc}

Using a simplified model for the flow geometry and magnetic field
structure in the head of a bow-shock pulsar wind nebula, we have
investigated the escape of high energy charged particles, and in
particular the role of the magnetic field in creating anisotropies in
the escaping flow. Following their
trajectories in the electric and magnetic field, we have try to
assess how much the escape probability depends on the magnetic field
geometry, in terms of strength of the field (the particle effective rigidity), the relative inclination
of the pulsar spin axis and pulsar kick velocity, and in particular we
focus on various kinds of anisotropies: {\it head to tail}, {\it
  inclination}, and especially {\it charge} anisotropy. 

We want to stress here that there are several other possible sources
of asymmetry in the escape of particles from BSPWNe: the energy flow
in the wind can have a  strong latitudinal dependence \citep{Spitkovsky06a,Tchekhovskoy_Philippov+16a} which will affect
the shape of the termination shock, and the structure of the flow
downstream of it, leading to the formation of fast channels \citep{Del-Zanna_Amato+04a}; particles
acceleration at the shock can depend strongly on the local conditions,
in term of inclination and magnetization \citep{Spitkovsky08a,Sironi_Spitkovsky09a,Sironi_Spitkovsky11a}, such that the shock itself
could introduce major asymmetries in the way particle are in injected;
shear at the CD between the internal pulsar wind  flow moving at a speed
$\sim 0.3-0.6 c$, and the slow moving
outer shocked  ISM (with velocity comparable to the pulsar kick
velocity), can produce 
Kelvin-Helmholtz instability which might
disrupt the magnetopause that is found at the CD in laminar models, and
introduce a further source of scattering and turbulence inside the
nebula. This will likely tend to isotropize the escape.

All of these effects can alter substantially the resulting escape
pattern, and they can also introduce major time
dependencies. However, even if the pattern of escape
particles can differ substantially from what we found, there are a few results from our study that we
deem robust:
\begin{itemize}
\item The fraction of escaping particles and the escape geometry is
  strongly energy dependent, and changing the energy from $\mathcal{R}
  =0.1$ to $\mathcal{R}=1$ can
  be enough to pass from an almost complete confinement, where most of
  the particles are advected in the tail, to an almost free escape. This
  means that energy independent arguments, in the description of how
  particles escape, are likely to be strongly inadequate, and only a
  detailed   modeling of the full trajectories can give trustable results.
\item Current sheets and lines are important confinement agents and
  they can strongly affect the pattern of escaping particles. This
  means that the structure, stability and dissipation of those
  features can substantially affect the way particles
  emerge out af these systems. In this respect the ability to properly
  model turbulence and dissipation, as key factors regulating those currents,
  becomes important for the level of anisotropy in the escape.
\item Particle escape is likely to be charge separated (strong charge
  asymmetry) or partially charge separated. If the high energy
  particles that escape carry a sizable fraction of the overall
  energetics, this means that BSPWNe are likely to be characterized
    by the presence of charge separated flows. Charge separated flows,
    (and the related return currents) can give rise to filamentation,
    Weibel, and two-stream instabilities, which might affect deeply the
    underlying MHD structure that is based on charge neutrality.
\end{itemize}

One of the main issue related to the escape of high energy particle is
that in systems like the Guitar nebula \citep{Hui_Becker07a} or the Lighthouse
nebula \citep{Pavan_Bordas+14a} one sided bright X-ray feature are observed, which
if interpreted in terms of high energy particles streaming out of the
nebula into the ISM magnetic field \citep{Bandiera08a}, require a large level
of anisotropy. Our results show that the required anisotropy can be due
to the internal magnetic field. Unlike reconnection, which only affects
low energy particles with Larmor radii smaller that the extent of the
reconnection site, which in general are much smaller than the size of
the bow-shock, global magnetic effects are relevant also for
$\mathcal{R}\sim 1$, and they generate large anisotropy over the
entire bow-shock head.

Even of greater importance is the fact that the escaping particles
might be in the form of a charge separated flow. This is highly
relevant in terms of self confinement of the escaping particles. As
the high energy particles start to stream in the ISM, they can drive
streaming instability in a fashion not dissimilar to that of cosmic rays
accelerated at SN shocks \citep{Ptuskin_Zirakashvili+08a,Malkov_Diamond+13a,Nava_Gabici+16a,DAngelo_Morlino+18a}. The level at which induced
turbulence saturate depends strongly on the level of charge
separation. In the presence of a net current, the magnetic field can be
amplified much more efficiently that for a neutral flow \citep{Skilling71a,Bell04a}. This means
that it is far easier to self confine the escaping particles in the
vicinity of the BSPWN, if the outflow is charge separated. Indeed the
recent TeV halo detected around Geminga \citep{Abeysekara_Albert+17a}, which has been
interpreted as an evidence for a larger ISM turbulence that what is
commonly assumed, could instead be due to local magnetic field
amplification by a charge separated pair outflow \citep{Evoli_Linden+18a}.

\section*{Acknowledgements}

The authors acknowledged support from the PRIN-MIUR project prot. 2015L5EE2Y "Multi-scale simulations of high-energy astrophysical plasmas".

\bibliography{Bib}{}
\bibliographystyle{mn2e}

\end{document}